\documentclass[11pt, a4paper,onecolumn]{article}

\usepackage{graphicx}
\usepackage{geometry}
\usepackage{amsmath}
\usepackage{hyperref}

\usepackage[default]{lato}

\usepackage{xcolor}
\usepackage{amssymb}
\usepackage[font=footnotesize,labelfont=bf]{caption}
\usepackage{graphicx}
\usepackage{natbib}
\usepackage{url}

\begin{document}

\twocolumn[
\vspace*{3cm}
{\LARGE Modular structure in labour networks reveals skill basins}\\

{\large
Neave O'Clery$^{1,2}$ and Stephen Kinsella$^3$ \\}

{ \footnotesize \it 
$^1$ Centre for Advanced Spatial Analysis, University College London, UK \\
$^2$ Alan Turing Institute, UK \\
$^3$ Kemmy Business School, University of Limerick, Ireland}

\vspace{0.3cm}
\noindent\rule{\linewidth}{0.35pt}
\vspace{0.3cm}

\setlength{\leftskip}{2.5cm}

A B S T R A C T

\setlength{\leftskip}{0cm}

\vspace{0.3cm}
\noindent\rule{\linewidth}{0.35pt}
\vspace{0.3cm}

\setlength{\leftskip}{2.5cm}

There is an emerging consensus in the literature that locally embedded capabilities and industrial know-how are key determinants of growth and diversification processes. In order to model these dynamics as a branching process, whereby industries grow as a function of the availability of related or relevant skills, industry networks are typically employed. These networks, sometimes referred to as industry spaces, describe the complex structure of the capability or skill overlap between industry pairs, measured here via inter-industry labour flows. Existing models typically deploy a local or 'nearest neighbour' approach to capture the size of the labour pool available to an industry in related sectors. This approach, however, ignores higher order interactions in the network, and the presence of industry clusters or groups of industries which exhibit high internal skill overlap. We argue that these clusters represent skill basins in which workers circulate and diffuse knowledge, and delineate the size of the skilled labour force available to an industry. By applying a multi-scale community detection algorithm to this network of flows, we identify industry clusters on a range of scales, from many small clusters to few large groupings. We construct a new variable, cluster employment, which captures the workforce available to an industry within its own cluster. Using UK data we show that this variable is predictive of industry-city employment growth and, exploiting the multi-scale nature of the industrial clusters detected, propose a methodology to uncover the optimal scale at which labour pooling operates.
\\
\\
{\it JEL Codes}: D85, D83, N9. \\
{\it Keywords}: Networks, knowledge flows, community detection, information diffusion.\\

\setlength{\leftskip}{0pt}

\vspace{0.3cm}
\noindent\rule{\linewidth}{0.35pt}
\vspace{0.3cm}
]
\onecolumn
\section*{Acknowledgments and funding}

We would like to acknowledge Eoin Flaherty, Mattie Landman and Camila Rangel Smith for research assistance on this project, as well as Daniel Straulino and Francesca Froy for their useful input.  

We would like to thank the Irish Department for Enterprise, Trade and Employment for their helpful feedback and engagement, as well as a number of colleagues for their insightful comments including Samuel Heroy, Helena Lenihan, Philip O'Regan, Martin Henning, Dario Diodato and Mercedes Delgado.  

This article was completed with support from the PEAK Urban programme, funded by UKRI's Global Challenge Research Fund, Grant Ref: ES/P011055/1, and Turing-HSBC-ONS Economic Data Science Award ‘Network modelling of the UK’s urban skill base’. 

\onecolumn
\section{\bf{Introduction}}

Emerging from the fields of economic complexity and evolutionary economic geography, a growing emphasis has been placed on the role of embedded knowledge in economic development processes \citep{nelson1982}. This literature emphasises the role of tacit know-how and skills embedded in workers, and in particular the development of individual specialisation \citep{jones2009burden, hausmann2011} which can be combined in firms and complementary teams \citep{neffke2019value} in order to drive economic growth and diversification \citep{hidalgo2007, hidalgo2018principle}.

From this perspective, the human capabilities embedded in a place constrain its economic and industrial development opportunities. This idea has been distilled into a general concept, that of 'relatedness' (see \citet{hidalgo2018principle} for a review). Specifically, places (countries, cities or regions) tend to growth or diversify into new industries and technologies that are 'related' to what already exists \citep{frenken2007theoretical, hidalgo2007}. Hence, places build on local capabilities in a path dependent manner. While there have been a wide variety of both theoretical and empirical proposals on which type of 'capabilities' drive these processes, recent work from \citet{coloc} suggests that skills and labour sharing increasingly dominate industry employment growth dynamics. 

In general, efforts to model these dynamics rely on metrics derived from industry networks with edge weights corresponding to some type of inter-industry relatedness metric (sometimes referred to as 'industry spaces') \citep{hidalgo2007}. These networks are used to build a variety of predictive metrics which aim to describe the potential of a place to grow or enter a new activity based on the presence of related activities. For example, the size or concentration of employment in 'related' sectors has been used to predict industry employment growth or new entry \citep{hidalgo2007, neffke2011, neffke2013skill, formality}. However, the vast majority of these metrics implicitly ignore the network structure and are based on local or 'nearest neighbour' links. If the goal is to the estimate the size or variety of the labour pool available to an industry, we need a metric that captures the inter-related set of industries that share skills and know-how in the broader neighbourhood (surrounding connectivity structure) of a node or industry. In order to do this, here we deploy tools from network science and connect with the literature on industry clusters in order to propose a new metric which captures the size of the 'skill-related' labour pool available to an individual industry. 

The role of labour and knowledge sharing within industrial clusters is well-known, and dates back to \citet{marshall1920economics}. Rather than geographical clusters, here we focus on industrial clusters based on skill-sharing which we uncover using data on inter-industry labour mobility. Such clusters promote firm learning and innovation \citep{march1991,catini2015}, and are key to innovation and labour pooling \citep{porter1998clusters, delgado2016defining}. We construct a network (industry space) of normalised inter-industry labour flows \citep{neffke2013skill}, and deploy community detection in order to uncover skill-related industry clusters on a range of scales. We can think of these scales as controlling the size of industry clusters or neighbourhoods in the network. We develop a new scale dependent metric,'cluster employment', which captures the size of employment in an industries' own cluster. In other words, cluster employment captures the number of workers with skills relevant to an industry. 

We show that this metric, using data from the UK, is associated with employment growth of industrial sectors. By letting the number of clusters vary from few large clusters to many small clusters, corresponding to a range of neighbourhood scales, we use a statistical fit criterion from our econometric model to uncover the optimal scale at which labour pooling operates. Our work contributes to a better understanding of labour markets and particularly segmentation of workers in distinct skill-based labour pools, and supports policy efforts around the generation of a class of more granular and targeted industrial and skills-focused education policies. 

The rest of this paper is organised as follows. In Section \ref{sec:lit} reviews the relevant literature, setting our contributions in context. In Section \ref{sec:data}, we introduce the data and methodology. In Section \ref{sec:results}, we present the results on industry clustering and the industry growth analysis. In Section \ref{sec:conclusionpolicy}, we conclude by discussing some of the policy implications and limitations of our work.

\section{Literature \label{sec:lit}}

\subsection{Modelling growth paths}

Regions are dependent on what they know. This is the fundamental tenet on which a large literature is based originating in the pioneering ideas of \citet{nelson1982}, and followed more recently by a surge in both conceptual and empirical contributions spearheaded by \citet{hidalgo2007} and \citet{frenken2007theoretical} amongst others. Much of this literature speaks of evolutionary 'branching' of economic activities \citep{frenken2007theoretical, essletzbichler2015relatedness}, whereby regions accumulate capabilities as workers learn on the job and then use this know-how to diversify into new economic activities in a path dependent manner. 

In order to model development paths for a region, it has been common to predict the entry of new industries or the growth of industry-specific employment using metrics based on the size or concentration of local employment in 'related industries' \citep{neffke2013skill, hausmann2021implied}. This type of metric captures the level of available relevant capabilities in the local economy. This `related diversification' literature has probed a wide range of questions around local growth paths, including employment and export growth (\citealp{hidalgo2007, hausmann2021implied}), firm and sector entry (\citealp{neffke2011regions, neffke2013skill}) and technological change (\citealp{boschma2015relatedness, rigby2015technological}). 

While some of these studies are agnostic as to the precise \textit{type} of capability overlap that connects related industries, either by design or lack of data, a growing literature aims to more precisely identify and measure different types of relatedness. In one example, a framework proposed by \citet{ellison2010causes} has enabled researchers to disentangle the relationship between a more general capability overlap measured via industry co-location and three distinct capability channels: customer-supplier sharing, knowledge sharing and labour pooling. Deploying this framework, using data on US manufacturing and services industries from 1910 to 2010, \citet{coloc} suggest that the customer/supplier channel has decreased in relative importance compared to the labour pooling channel over time, and that the labour channel is particularly potent for services industries. \citet{o2021unravelling} extended this work to look at the relative importance of each channel for industry clusters, showing that co-location patterns of complex services are more heavily associated with skill-based linkages. 

Investigating employment growth and diversification patterns, \citet{rossi2007establishment} link employment growth to the accumulation of industry-specific human capital. In related work \citet{neffke2011regions},  \citet{boschma2013emergence} and \citet{essletzbichler2015relatedness} show that firms are more likely to diversify into industries that are linked to core activities via skill-based ties than industries that are linked by value chain linkages, and \citet{jara2018role} show that pioneers firms (i.e., the first firm in an industry in a region) are more likely to survive and grow if their workers had experience in related industries rather than related occupations. Overall, the literature suggests that a key factor driving the growth and diversification of industries in a region is the local availability of relevant know-how in the form of skills learned on the job in related industries. 

We investigate the dynamics underlying the growth of employment in both the full set of industries and services industries. As argued by \citet{coloc}, services are typically more difficult to trade than manufactured goods and will hence need to locate alongside their customers. With respect to labour, services tend to employ skilled workers, and place a premium on face-to-face interaction with customers compared to manufacturing industries. Hence, access to experienced and appropriately skilled workers is particularly important in this case. 

\subsection{Industry spaces}

A suite of network based models to describe the process by which places build on local capabilities to move into new economic activities in a path dependent manner have been developed, the most well known of which is the Product Space of \citet{hidalgo2007}. In general, in such models, nodes represent industries, products or technologies, and edges represent capability- or skill-overlap \citep{hidalgo2007, neffke2011}. This approach aims to model industry diversification and growth as a dynamical process on a network, akin to a diffusion/spreading or random walker process, whereby a region can sequentially jump into new industries based on probabilities defined by edge weights. The network can be seen as an underlying 'landscape' or map upon which the dynamics occur. The topology or global structure of this network determines the dynamics for a particular initial condition (i.e., the starting subgraph of industries present in the region).

A wide variety of approaches have been deployed to estimate the edge weights of this network, each with a slightly different meaning or interpretation. A well-known example, the Product Space, deploys geographic co-export of products as a general metric for capability overlap \citep{hidalgo2007}. Other approaches aim to capture a more specific type of capability overlap, for example input-output tables can be deployed as a proxy for customer-supplier sharing \citep{acemoglu2015networks} and the co-appearance of industry pairs on patents can be used as a proxy for knowledge sharing \citep{ellison2010causes, jaffe1989characterizing}. A third type of capability overlap involves labour and skill sharing, which we focus on here. This can be measured via co-production of goods in plants \citep{neffke2011regions}, occupational similarity \citep{farjoun1994beyond, chang1996evolutionary} or job switches \citep{neffke2013skill}.

Focusing on skill-proximity between sectors, we adopt the latter approach and build our industry network using the number of workers who switch jobs between industry pairs as edge weights \citep{neffke2013skill}. Intuitively, if many workers move from one industry to another, then it is likely that these industries share a high degree of skill-similarity. This approach is less noisy than alternative metrics, and has comparable estimates across all sectors. Industry networks based on inter-industry labour flows have been deployed to study a number of phenomena, including the growth of formal organised employment in developing cities \citep{formality, o2019commuting}, labour market resilience \citep{diodato2014}, affirmative action labour market policies \citep{landman2020impact} and FDI spillovers \citep{csafordi2020productivity}. 

Despite widespread use of industry spaces to model diversification and growth paths, the vast majority of existing metrics implicitly ignore the network structure and are based on local or 'nearest neighbour' links, and thus ignore the wider network topology. Hence, only immediate connections are considered relevant, and the complex inter-connection structure in the wider neighbourhood of the node beyond first order neighbours is neglected. In essence, currently modelling approaches are at odds with the aforementioned theoretical narrative concerning evolutionary 'branching' of economic activities \citep{frenken2007theoretical, essletzbichler2015relatedness}, and broad empirical use of networks as a landscape on which to model growth and diversification processes. Limited previous work has looked at diffusion dynamics on product networks, aiming to model diversification paths over time \citep{hidalgo2007} and diversification strategies \citep{alshamsi2018optimal}. 

Here we take a distinct approach which enables us to connect to the literature on industrial clusters. Specifically, we propose that meso scale structures in the network, which correspond to node communities, describe the neighbourhood of a node in the network through the prism of industry clusters. This approach delimits the skills and capabilities available to an industry, and provides a natural framework in which to model development paths which captures the broader network structure around a node.  

\subsection{Industrial clusters \label{ind_clusters}}

The study of industrial agglomeration patterns is connected to work on industrial clusters \citep{porter1998clusters, porter2003economic} whereby geographically co-located groups of firms generate positive spillovers via sharing of various costs, and reap other benefits such as competition which drives productivity \citep{porter2011competitive}, and local demand effects \citep{Fujiteetal2001}. Clusters are an important policy tool, and form a key tenet of the EU's 'smart specialisation' strategy \citep{boschma2017relatedness}.

Here we propose a methodology to uncover skill (rather than geography) based industrial clusters, which we term 'skill basins'. Specifically, industry clusters detected in labour networks correspond to groups of industries which exhibit high internal mobility and low external (inter-cluster) mobility. Hence, these are groups of industries which exchange labour, skills and know-how, and thus illuminate both labour mobility patterns within an economy and enable us to delineate the relevant skills available to an industry. 

The presence of skill-based industrial clusters poses clear benefits to workers and firms. As mentioned above, industries (and particularly service industries) frequently co-locate in order to benefit from shared labour pools, accessing skilled workers and sharing costs associated with search and matching processes \citep{marshall1920economics, ellison2010causes}. Clusters defined by labour mobility patterns further provide information on segmentation of the labour force \citep{wilkinson2013dynamics}, and illuminate knowledge flows between firms which support learning and innovation \citep{march1991} (see \citet{iammarino2006structure} for a discussion of knowledge transfer within industrial clusters). We note that knowledge flows may be achieved via a variety of mechanisms including but not exclusive to labour mobility, including R\&D collaboration and patenting.

On the other hand, the formation of skill clusters may also pose risks, limiting worker and knowledge flow between different sectors. Adopting an evolutionary perspective on economic resilience, which takes a non-equilibrium view of resilience as the ability of a region or place to adapt to a shock rather than return to a pre-shock equilibrium \citep{boschma2015}, it is clear that clusters may impede the ability of firms and workers to move between sectors in response to adverse events. Specifically, \citet{diodato2014}, \citet{eriksson2016industrial} and \citet{straulino2021res} find that inter-sectoral mobility is an important factor for regional recovery from an employment shock. More mobile workers, with more opportunities for transition between sectors, are better protected in the event of a crisis or a reorganisation of work.

Another well-known feature, competition, has long been seen as a benefit of cluster formation \citep{porter2003competitive, porter2011competitive}. In particular, it is thought to foster the pursuit and rapid adoption of innovation. But its statistical effect on productivity and employment in the empirical literature is mixed. For example, using a French dataset, \citet{combes2000economic} found that competition had a negative impact on employment growth. Using Canadian data, however, \citet{wang2017industry} found a positive effect. 

Here we uncover industry clusters from the analysis of inter-industry labour flows in a network. These clusters are interesting and informative in their own right, but also govern dynamics on the network. Harnessing this insight, we use these clusters to modify a traditional (nearest neighbour) metric, aimed at capturing employment in related industries, to incorporate information on the network structure in the broader neighbourhood around the node.

In previous related work, focused on detecting clusters in similar networks, \citet{delgado2016defining} has sought to identify industry clusters for US industries using inter-industry linkages based on co-location patterns, input-output links, and similarities in employment shares across occupations. The latter occupational similarities aim to capture skill-overlap in much the same manner as labour flows, but are thought to be significantly less precise \citep{neffke2013skill} and do not quantify labour mobility as such. In further related work to ours, \citet{park2019global} use LinkedIn data to construct a global labour flow network of firms, and apply community detection techniques to uncover firm clusters.

\subsection{Network structure and dynamics}
 
Network analysis provides a uniquely powerful tool to understand and quantify complex systems whose aggregate dynamics depends not on individual agents or homogeneous populations but an underlying heterogeneous interconnection structure. Network models are increasingly used to understand the role interconnection structures play in economic and innovation-related processes, including research clusters \citep{catini2015}, innovations \citep{hermans2013niches}, worker skill complementarity \citep{neffke2019value}, country-level R\&D efficiency \citep{guan2012modeling}, and the success of venture capital markets \citep{milosevic2018skills}. Of particular relevance to this work are studies related to industry-networks, including regional skill relatedness (\citealp{fitjar2017regional, neffke2013skill}), and the inter-industry propagation of microeconomic shocks to macroeconomic outcomes (\citealp{acemoglu2015systemic, gabaix2011granular}). 

Here we are concerned with a network of industries, where the edge weights correspond to the number of workers who transitioned between the industry pair during a certain period. In practice, following \citet{neffke2018}, we normalise this count with respect to the number of workers who would have moved at random given the respective size of each industry (similar to the Configuration Model of \citealp{molloy1995critical}). Hence, the eventual edge weight is the 'excess' movement relative to that expected at random. 

While there are a wide range of tools and approaches to studying network structure \citep{newman2003structure}, we will focus on uncovering modular structure. The presence of densely connected communities of nodes, with sparse connections between communities, is indicative of an underlying sub-structure or functional organisation \citep{fortunato2010community}. Community detection has been used extensively to study the structure and dynamics of biological and social networks \citep{girvan2002community}. While most well-known methods for community detection seek to find a single node partition under a particular optimisation strategy (e.g., modularity), it is more natural to think about a range of partitions on different scales (from many small node clusters to few larger clusters). This information can be extracted by analysing the patterns of random walkers on a network: walkers tend to get trapped in densely connected regions. The Stability method \citep{delvenne2010stability} uses a resolution parameter to control the time the walkers can roam. Shorter time scales (lower resolutions) allow for less exploration of the network resulting in smaller communities, while longer time-scales (high resolutions) correspond to larger node aggregations. Using this approach, we extract a hierarchy of partitions of industries into clusters corresponding to different scales. This can been seen as a kind of alternative multi-level industry classification, where each level corresponds to a particular strength of clustering (scale or resolution).

We specifically chose this approach to community detection as it is based on a spreading process on the network, akin to modeling path dependent diversification as a multi-step diffusion process. We can interpret the partition found at resolution $\tau$ to be the set of nodes reached by a spreading process in $\tau$ steps. Hence, by varying the resolution, we vary the size of neighbourhood around the node that we consider in our model.

We use this feature to develop a network-based metric to predict industry-region employment growth. Instead of estimating the size of the skilled workforce that is locally available to an industry via the standard approach outlined above (which implicitly ignores the network structure), we develop a new metric based on the total employment within an industry's own cluster, denoted 'cluster employment', and is based on the idea that our clusters naturally delineate the size and industry composition of the skilled labour pool available to any industry. 

Capturing a nodes' neighbourhood or cluster is important in a number of ways beyond those outlined above. Firstly, the cluster captures a broader swathe of workers with relevant skills for the industry, even if accessed via multi-step jumps. Secondly, the neighbourhood size is bespoke for each node, and depends on the degree of connectivity around the node. Hence, we allow for heterogeneity in terms of the reach of nodes into the network, meaning that nodes in large dense parts of the network will have access to much larger labour pools while those in isolated clusters will be much more limited. Thirdly, any measurement noise in the network (e.g., missing or mis-estimated edges) will be lessened by taking into account clusters rather than just neighbours.

We exploit the information produced by our multi-resolution community detection analysis to vary the size of the neighbourhood based on strength of connectivity or spreading time, which enables us to reveal the statistically 'optimal' scale at which labour pooling operates within a given context. We do this by probing the relationship between industry employment growth patterns and our 'cluster employment' metric across a range of scales, and inferring the optimal scale from the statistical strength of this relationship.

\section{\bf{Data and Methods}}
\label{sec:data}
\subsection{\it Data}

We use the UK Annual Survey of Hours and Earnings dataset (ASHE) in order to construct the inter-industry flow matrix. This dataset enables us to track the industry code of a large cohort of UK employees over the period 2009 to 2018. Specifically, it contains anonymised demographic and employment information of 1\% of the total employee jobs in the HM Revenue \& Customs (HMRC) Pay As You Earn (PAYE) records.

A longitudinal dataset, it follows a large cohort of workers over time. For every worker, the dataset includes information on various variables, from individual characteristics, such as age and sex, to employment information, including pay, occupation and industry. The sample does not include the self-employed. We construct our flow matrix for the maximum period 2009-2018 due to the sparsity of the sample.

To compute total employment by industry, city and year, we use the Business Register and Employment Survey (BRES), which contains employment records from all registered firms in the UK\footnote{We do not include Northern Ireland, and hence technically our cities correspond to Great Britain.}. We use years 2010-2015. Both ASHE and BRES are compiled by the Office of National Statistics (ONS) and are accessed under strict confidentiality agreements and arrangements. 

We use functional urban areas (FUAs) as defined by the OECD as our city definition (there are 105 FUAs in Great Britain). Industries correspond to the 4 digit level standard industrial classification (SIC 2007). 

\subsection{\it Network construction}

We construct the labour network based on the skill-relatedness methodology of \citet{neffke2018}. First, we calculate skill relatedness for each year. The number of employees who transitioned between two industries $i$ and $j$ between years $t$ and $t+1$ are denoted by $F_{ijt}$. The skill-relatedness is expressed as: 
$$
SR_{ijt}=\frac{F_{ijt}}{F_{jt} F_{it}/F_t}
$$
where missing indices mean all values are included in the variable. The denominator represents the worker flow between industries $i$ and $j$ that would be expected at random given the total flows of the respective industries. This is known as the Configuration Model \citep{molloy1995critical} in the network science literature. Hence, the skill relatedness metric (edge-weight) captures excess flows beyond what would be expected at random.

The skill relatedness measure is highly skewed, with industries that are more related than expected ranging from 1 to infinity and those that are less related than expected lying between zero and one. Therefore, following \citet{neffke2018}, we transform it so that it maps onto the interval [-1, 1):
$$
\tilde{SR}_{ijt}=\frac{SR_{ijt}-1}{SR_{ijt}+1}.
$$
To improve the precision of the indicator and protect anonymity, we average it across all yearly flows between 2009 and 2018:
$$
M\tilde{SR}_{ij}=\frac{1}{10}\sum_{t=2009:2018} \tilde{SR}_{ijt}
$$
and make it symmetric:
$$
S\tilde{SR}_{ij}=\frac{M\tilde{SR}_{ij}+M\tilde{SR}_{ji}}{2}.
$$
We consider the weighted undirected adjacency matrix:
\[ A_{ij}^\gamma =
  \begin{cases}
    S\tilde{SR}_{ij}       & \quad \text{if } S\tilde{SR}_{ij}>\gamma\\
    0  & \quad \text{otherwise}
  \end{cases}
\]
Setting $\gamma=0$ corresponds to keeping flows for which $SR_{ijt}>1$, and hence the proportion of flows is larger than would be expected at random.  

\subsection{\it Community detection}

Our aim is to detect groups of industries that exhibit high internal worker mobility in the form of network communities. There are a wide array of algorithms and techniques to perform this task as outlined above. Most community detection algorithms are composed of two components: an optimization criteria which is minimized (or maximized) by a particular partition of the nodes, and a searching algorithm which looks for the best node partition to satisfy the optimization criteria. 

Here, we deploy an optimization criteria based on simple random walk model, which is called Stability (\citealp{delvenne2010stability, lambiotte2008dynamics, lambiotte2011flow}). The core idea is that if a random walker - who jumps from node to node with probability proportional to the edge weights - gets trapped in a region of the network (set of nodes) for a prolonged period of time this indicates a region of densely connected nodes which form a community. The Stability optimization criteria, elaborated on in Section 2 of the Appendix, seeks to maximize the probability of a node remaining within a community for a time period, denoted by resolution parameter $\tau$.

One of the key advantages of this approach is that it detects communities on a range of scales, ranging from a large number of small communities to few large communities. Intuitively, if we let a random walker wander for longer periods on the network, the walker will encounter larger and larger communities. Hence, the parameter $\tau$ corresponds to a resolution or scale. For smaller values of $\tau$, we detect many small communities, while for larger values of $\tau$, we detect fewer, mostly larger communities. While there are other algorithms which can detect communities corresponding to different resolutions, including an adapted version of modularity which includes a resolution parameter \citep{reichardt2006statistical}, the Stability approach connects the resolution parameter to an intrinsic notion of time within a dynamical framework, which is more appropriate for our application. More broadly, Stability has a mathematical relationship to several well known community detection optimizations. It can be seen as a generalisation of modularity \citep{newman2004finding} and Normalised Cut \citep{shi2000normalized}, which corresponds to $\tau=1$, and Fiedler's spectral method \citep{fiedler1973algebraic} which corresponds to $\tau=\infty$.

The second component, the searching algorithm, looks for the best node partition to satisfy the optimization criteria. This latter problem is NP-hard, and we therefore use heuristic methods, such as Louvain's algorithm \citep{blondel2008fast}, to solve it. 

More detail on these algorithms is provided in the appendix.

\subsection{\it A network-based model for industry employment growth}

Many previous studies have used industry networks to model the process by which regions or countries move into 'related' industries (see \citet{hidalgo2018principle} for a review). These studies (e.g., \citealp{hausmann2021implied,coloc}) typically estimate relationships of the form:
$$
G_{ir}=\alpha + \beta_0 E^0_{ir}+\beta_1 RE^\gamma_{ir} + \psi_i+\eta_r +\epsilon
$$
where $E^0_{ir}$ is the employment in industry $i$ and region $r$ at time $t_0$, and $G_{ir}$ is the growth in employment between times $t_0$ and $t_1$:
$$
G_{ir}=\log{E^1_{ir}}-\log{E^0_{ir}}.
$$
and $\psi_i$ and $\eta_r$ are industry and region fixed effects. The expression $RE^{\gamma}_{ir}$ is the 'related employment', or the size of employment in proximate industries in the network (in the region) at time $t_0$:
$$
RE^{\gamma}_{ir}=\sum_{j\neq i} \frac{A^{\gamma}_{ij}}{\sum_{k\neq i} A^{\gamma}_{ik}} E^0_{jr}.
$$
Hence, for a region $r$, this is the sum of employment in all industries (except $i$) weighted by their (normalised) edge weight to node $i$. The related employment can be seen as the size of the potential labour pool with relevant skills or capabilities. For a positive and significant coefficient $\beta_1$, we can infer that a larger pool of similar skills for industry $i$ (in region $r$) is correlated with a higher rate of employment growth for industry $i$. 

The expression $RE^\gamma_{ir}$, however, is a local metric in the sense it does not take into account the network structure other than weight the employment in neighbouring industries by the normalised edge weights. We propose an alternative form based on the community structure, whereby industries have access to labour only within their own community, which we refer to as the 'cluster employment': 
\begin{equation}
\label{eq_ce}
CE^{\tau,\gamma}_{ir}=\sum_{j\neq i \cap j \in C^\tau_i} \frac{A^\gamma_{ij}}{\sum_{k\neq i \cap k \in C^\tau_i} A^\gamma_{ik}} E^0_{jr}
\end{equation}
where $C^\tau_i$ denotes the set of nodes in the community of node $i$ at resolution $\tau$. This approach captures the size of the local skilled workforce available to an industry accounting for the modular nature of the network structure. In general, we estimate a model of the form 
\begin{equation}
\label{main}
G_{i,r}=\alpha + \beta_0 E^0_{ir}+\beta_x XE^{\tau,\gamma}_{ir} + \psi_i+\eta_r + \epsilon
\end{equation}
where $XE^{\tau,\gamma}_{ir}=CE^{\tau,\gamma}_{ir}$ or $RE^{\gamma}_{ir}$. 

\begin{figure*}[t!]
\centering
\includegraphics[width=5.8in]{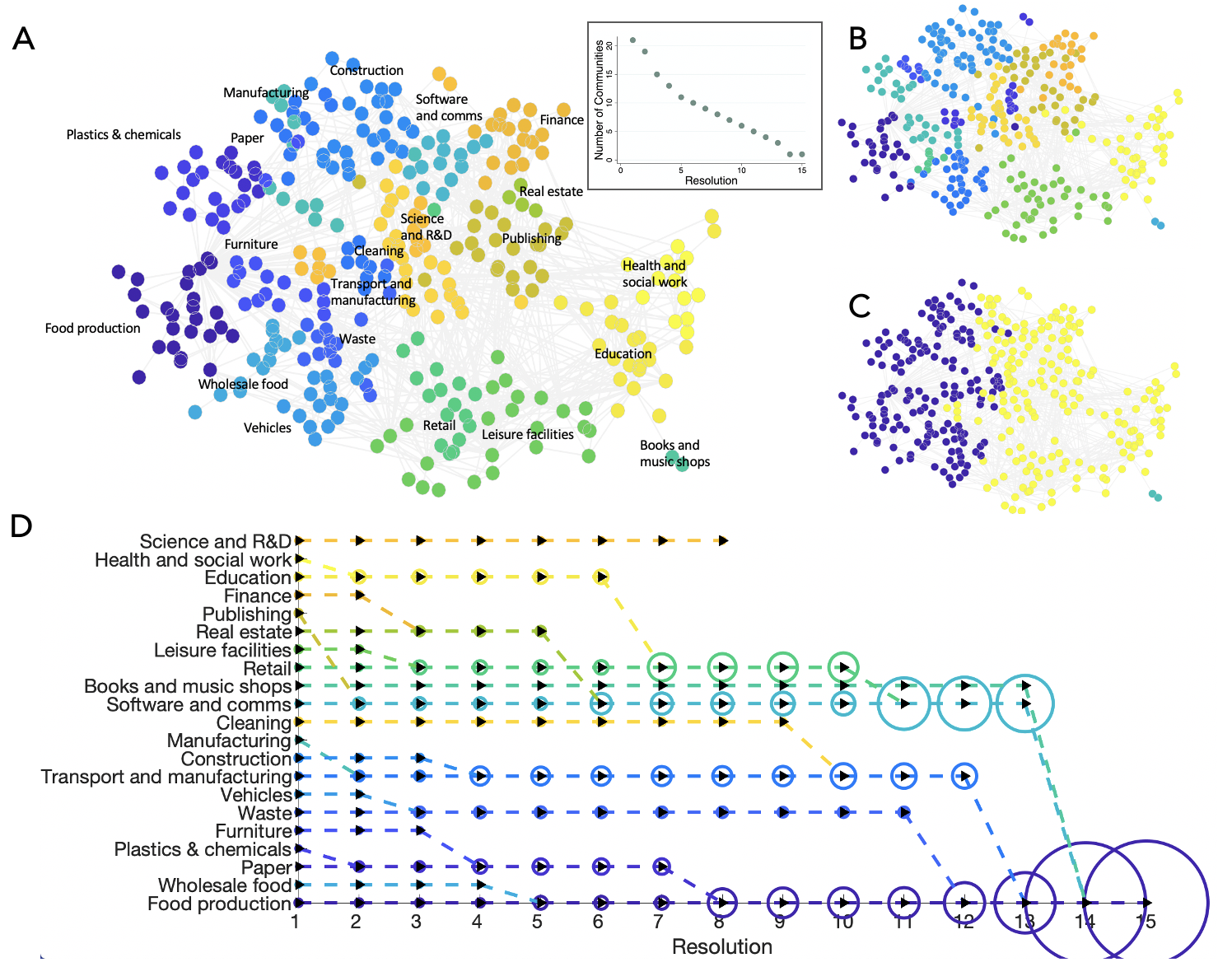}
\caption{A. The UK labour flow network, with nodes coloured by community (resolution $\tau=1$). The node layout is based on a spring algorithm called 'Force Atlas' in Gephi. The inset shows the number of communities for each value of $\tau$. As the resolution time increases, the number of communities decreases. B/C. The network with nodes coloured by community at resolution $\tau=4$ and $\tau=13$. D. A dendrogram illustrates how nodes are grouped for all values of $\tau$. 
\label{fig1}}
\end{figure*}

Due to the 1\% sample used to construct the UK adjacency matrix, which results in a sparse dataset, we use an adjacency matrix built from years $2009-2018$ to predict employment growth in the UK from $t_0=2010$ to $t_1=2015$. Although we do not have an ideal out-of sample estimation (following \citealp{hausmann2021implied, coloc}), the structure of this type of adjacency matrix is known not to change significantly over relatively short periods of time \citep{neffke2017inter} and hence this is highly unlikely to materially affect any of our results. 

We use a subset of industries that form the largest connected component of the labour flow network, omitting a small number of peripheral industries which experienced no labour flows (job transitions) with other industries within our time period. This leaves us with 339 industries (SIC 2007).

Following the literature, for the majority of the analysis below, we set $\gamma=0$. We perform a sensitivity analysis, however, in the appendix to check to what extent the choice of $\gamma$ affects our key results.

\section{\bf{Results}}
\label{sec:results}

\subsection{\it Industry clusters}

First, we focus on exploring the multi-scale modular structure of the UK inter-industry labour flow network. Beyond forming the basis for our new metric, insights from this preliminary analysis are key for place-based policy-making as discussed in Section 5.1 below. 

Figure \ref{fig1} A shows a visualisation of the labour flow network for the UK. The node layout is based on a spring algorithm called `Force Atlas' in Gephi\footnote{Gephi version 0.8 was used.}. Edges are shown over a threshold $\gamma=0$. We observe a large degree of clustering of related industries, with public services broadly located on the far right-hand side, retail leisure on the bottom right and finance and professional activities on the top right. Manufacturing and related industries appear slightly less clustered and dominate the left, with construction on the top left and food and farming on the bottom left. 

In order to systematically extract industry groupings corresponding to labour mobility and skill sharing patterns, we apply a community detection algorithm as introduced above. A key feature of this algorithm is that it produces not one, but several node partitions corresponding to industry clusters at different scales. 

The inset of Figure \ref{fig1} A shows the number of communities as the resolution parameter increases. As the parameter increases, the community detection algorithm finds increasingly larger industry clusters, effectively merging smaller clusters into large industry groupings\footnote{The algorithm is not strictly hierarchical in the sense that the clusters of partition $k$ are not necessarily nested in partition $k+1$.} Figure \ref{fig1} B and C shows the communities at resolutions $\tau=4$, and 13. 

A couple of features are evident at first glance. For example, by $\tau=13$ services including the public sector and finance have merged with business and software activities (yellow cluster), but remain distinct from retail, farming, manufacturing and construction (blue cluster). This highlights a clear segmentation of the economy, whereby workers and skills rarely transition from services to manufacturing and vice versa. This is a particularly striking finding which is consistent with the well established view that large swathes of traditional 'blue collar' workers are being left behind in the `knowledge economy'.   

We wish to examine in more detail the structure of the partitions found as the resolution parameter increases. Specifically, if a node community persists  over a range of resolutions or scales, as other communities merge into larger groupings, then we can deduce that this community exhibits particularly high internal connectivity and weak connections to other clusters. On the other hand, if a node community merges quickly at a relatively low value of the resolution parameter, then this cluster enjoys stronger connections to other communities in the network. A dendrogram, seen in Figure \ref{fig1} D, quantifies this merging process\footnote{The partitions generated by the community detection algorithm are not strictly nested. A simple majority rule is deployed here to produce the dendrogram.}. We observe that eventually, as the resolution parameter increases, all clusters are merged into a large single cluster by $\tau=14$.  

On short timescales, we observe merging on the services side between health, social work and education as well as finance and real estate. We also observe a number of fast merges in manufacturing food-based industries. These early merges correspond to initial groupings which exhibit tight connections to other clusters. Many of these new groupings remain intact until they merge with other groups later. For example, finance and real estate merges with software and comms at $\tau=6$, and health, social work and education merges with retail at $\tau=7$. A large number of clusters also remain unmerged until high values of $\tau$. These are tightly knit clusters with few strong connections to other clusters. These include, for example, waste and cleaning. By $\tau=11$, a super-cluster of services industries has formed, while most manufacturing and heavy industries have merged by $\tau=13$. Eventually, at $\tau=14$, both large clusters (as well as books and music) are forced to merge into a single cluster. 

This analysis highlights the heterogeneous nature of industry clusters based skill-sharing. As opposed to official sectoral groupings, some industries cluster into tightly knit groups of very similar activities, rarely exchanging workers with other sectors. Other industries are connected to a diverse set of other industries, exhibiting a large number of connections and flows. These groupings ebb and flow as the resolution parameter changes, implicitly tuning the degree of connectivity required for cluster formation. This motivates us to consider a bespoke 'neighbourhood' for each node when estimating the size of its labour pool, with neighbourhood size dependent on the resolution selected, as captured by our cluster employment variable.  

\begin{table*}[t!]
\centering
\includegraphics[width=0.6\linewidth]{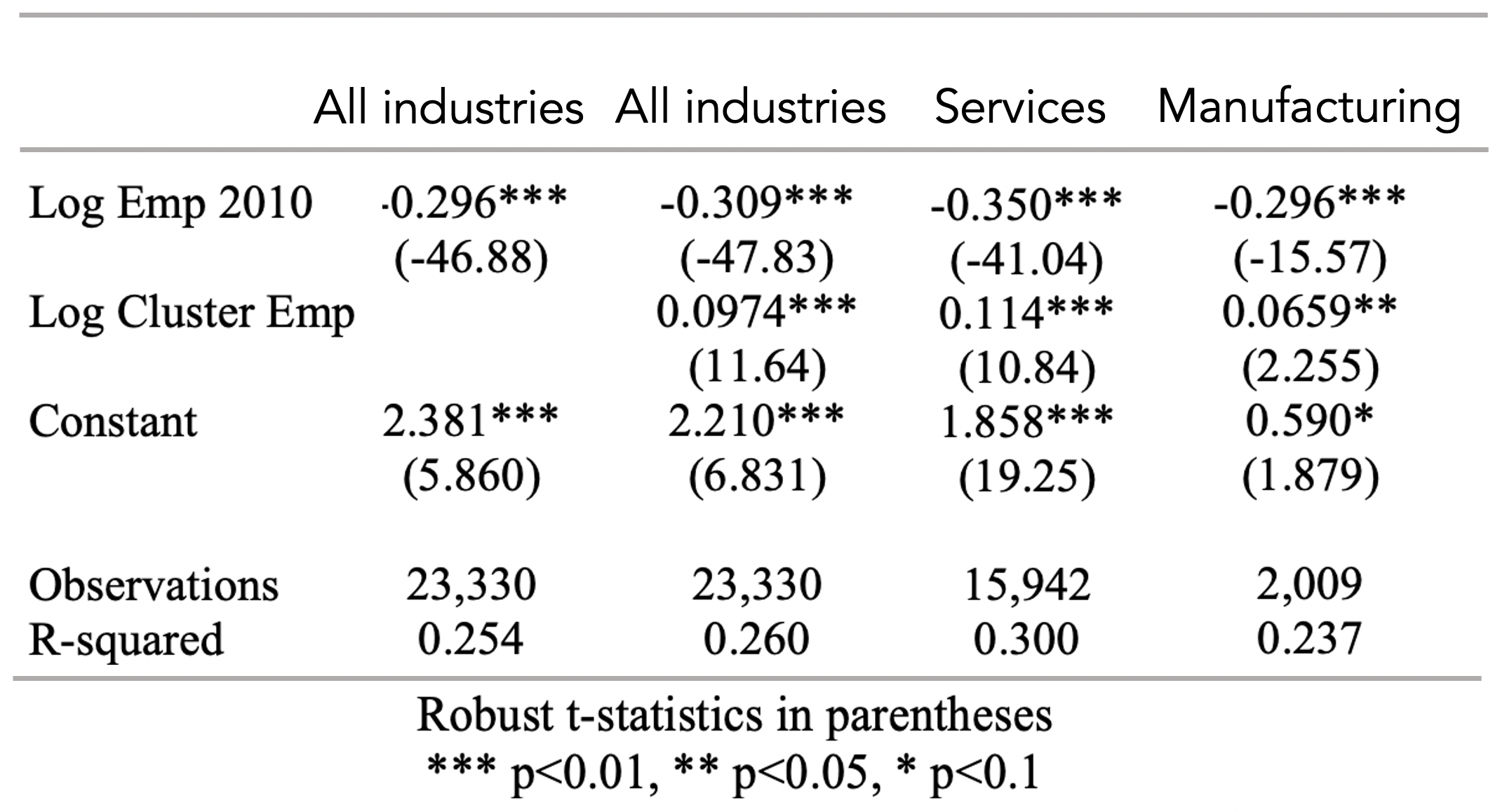}
\caption{OLS for industry employment growth as a function of log employment and cluster employment (CE) in the base year as defined by Eqn \ref{main} - for all industries, manufacturing only and services only - including city and industry fixed effects. We use an edge threshold of $\gamma=0$ and resolution $\tau=4$. \label{fig2}}
\end{table*}
\subsection{\it Uncovering the scale at which labour pooling operates}

Following a rich literature (cited extensively above) that supports the theory that industrial diversification and growth processes depend on the availability of relevant skills and capabilities, we wish to explore the relationship between 'cluster employment', capturing the size of the labour pool of related skills available to industries, and industry growth patterns. Specifically, we wish to investigate if a particular scale (or division of industries into clusters of a characteristic size) is more predictive of growth patterns. If we find the model performs best for coarse partitions with few clusters each with many industries, then it implies that industries look deep into the network, and tap distant industries for new workers and skills. If we find the model performs best for fine partitions (many clusters each with few industries), then it implies that industries source skills only from very similar industries. 

By definition, cluster employment depends on a chosen partition. For ease of notation, we drop the $\gamma$ from CE and use CE$^\tau$ to refer to cluster employment computed for resolution $\tau$\footnote{Following previous literature, here we use an edge threshold of $\gamma=0$. A sensitivity analysis can be found in the Appendix.}. 

First, we investigate the relationship between cluster employment and industry employment growth for single partition. Table \ref{fig2} shows the OLS for industry employment growth as a function of log employment and cluster employment in the base year for a chosen partition. Considering the full set of industries and services alone, we observe a statistically significant relationship between industry-city employment growth and cluster employment. The coefficient of the size of employment in the base year is a negative, which is also consistent with the literature and due to mean reversion effects. 

Next, we split the set of industries into services and manufacturing. We find a more significant association in the case of services which corresponds to previous results showing that service employment growth is more strongly related to labour availability in related industries compared to manufacturing sectors \citep{coloc}. In general, the size and sign of the coefficients and statistical indicators are consistent with the literature. 

Our main goal, however, is to deploy our framework to estimate at which scale does labour pooling operate when it comes to supplying growing industries with suitably skilled labour. In other words, what is the statistically 'optimal' partition of the network into groups of industries that share 'sufficiently related' workers and skills to assist industry growth? 

\begin{figure*}[t!]
\centering
\includegraphics[width=\linewidth]{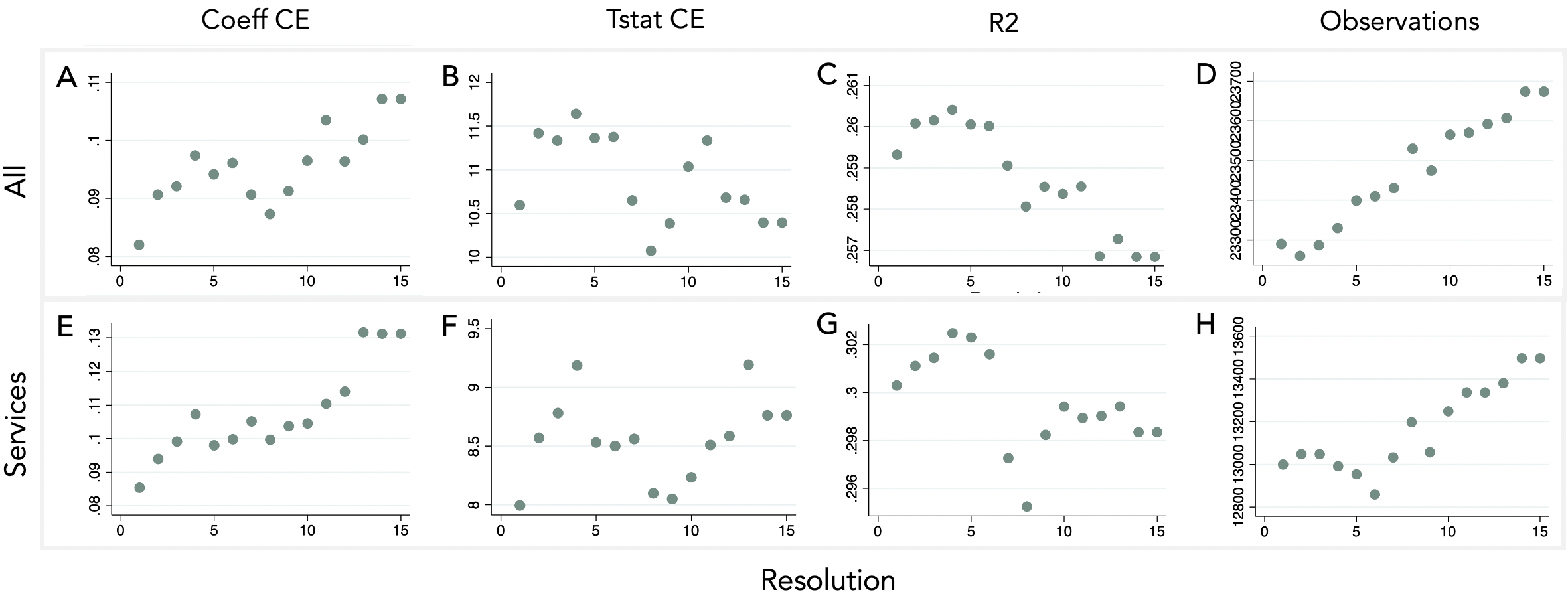}
\caption{The coefficient of CE (column 1), tstat (column 2), R2 (column 3) and number of observations (column 4) corresponding to column 2 (all industries) and 3 (services) of Table \ref{fig3} for variation in the cluster resolution $\tau$. \label{fig3}}
\end{figure*}

In order to investigate this we repeat the model from column 2 and 3 of Table \ref{fig2}, computing CE$^\tau$ for all values of $\tau$. As $\tau$ increases, the size of the clusters increases and industries can 'reach' further in the network in order to access more distant related skills. We plot the coefficient and t-statistic of CE$^\tau$, and r-squared and number of observations in Figure \ref{fig3}, showing results for both the set of all industries and services. We focus on the r-squared as a measure of statistical fit. We observe a peak in the size of the r-squared around $\tau=4$ for both the full set of industries and services.

A detailed visual representation of the partition corresponding to $\tau=4$ is provided in Figure \ref{fig1} B. It is composed of 13 communities, each containing between 2 and 54 industries (average size is 26 industries). We can clearly see that some key merges have already occurred by $\tau=4$. On the services side, we have  health and education, retail and leisure, finance and real estate, publishing and software/comms. On the manufacturing and heavy industry side, we have vehicles and waste, paper, furniture, plastics and chemicals. These groupings can be seen as integrated labour pools, frequently sharing workers, skills and knowhow. 

However, what is clear from Figure \ref{fig3} column 4 is that the number of observations is changing depending on the value of the resolution parameter $\tau$. This occurs because for each $\tau$ there exist some singleton clusters containing just one industry. Cluster employment is not defined for such industries as there is no skill-related employment pool available to them under our definition, and these observations are dropped. For finer divisions of the network (corresponding to lower values of $\tau$) there exist more of these. Singletons are typically peripheral industries with no strong connections to any cluster. Hence, the peaks in predictive power seen in Figure \ref{fig3} for specific values of $\tau$ may be as a result of dropping the 'right' industries - those that are peripheral in the network and thus likely smaller in terms of employment size and more difficult to predict - rather than any information bestowed by the non-singleton industry partition itself. 

\begin{figure*}[t!]
\centering
\includegraphics[width=\linewidth]{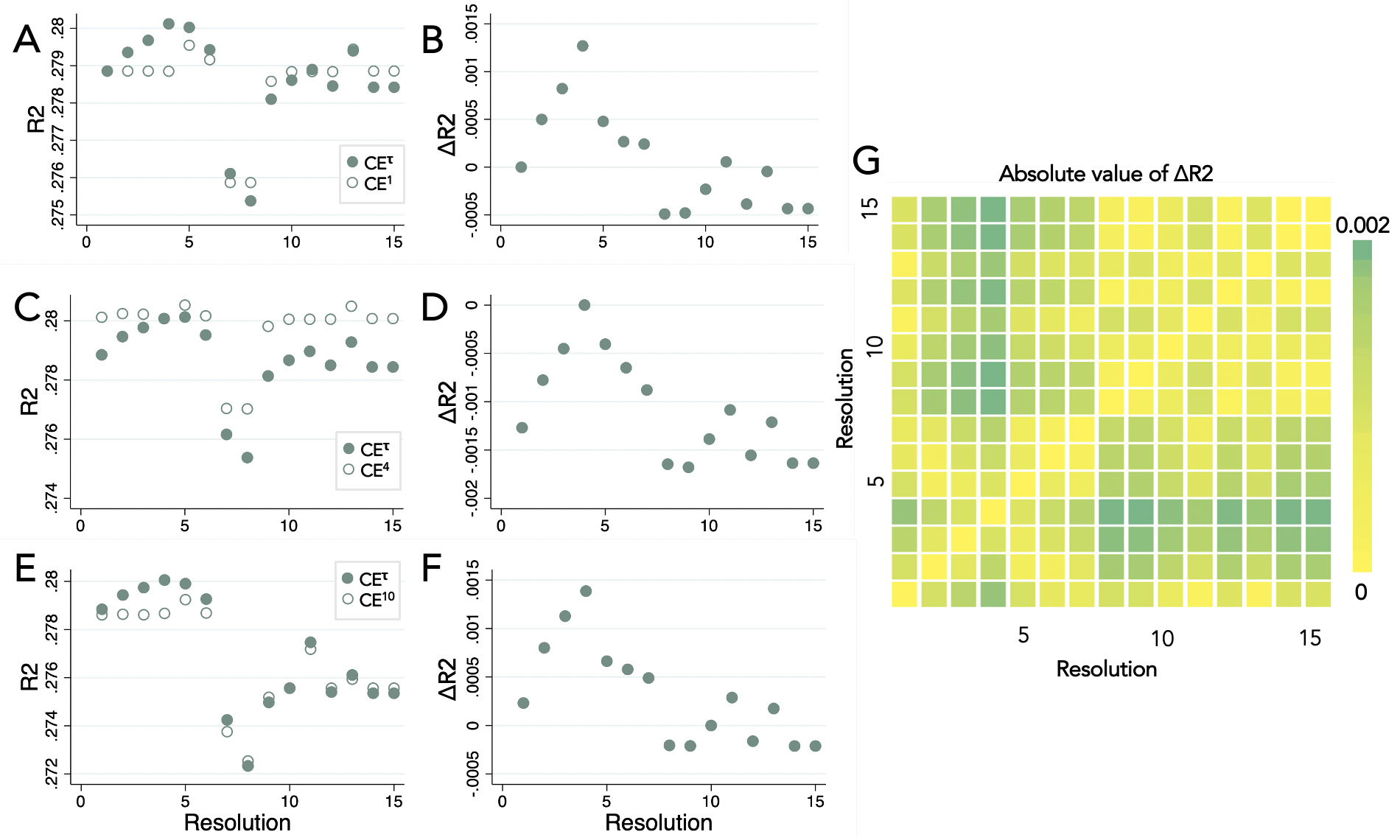}
\caption{A-F. We fix the number of observations at each resolution, and compare the r-squared of CE$^\tau$ defined as above to the r-squared of CE$^1$ (many small clusters), CE$^4$ and CE$^{10}$ (few large clusters). G. Absolute value of $\Delta$R2 (difference in r-squared) between pairs of CE$^\tau$ with identical number of observations. Rows or columns of darker entries indicate values of $\tau$ for which the difference in r-squared is large. \label{fig4}}
\end{figure*}

In order to investigate this question, we fix the number of observations at each resolution, and compare the r-squared of CE$^\tau$ for two different values of $\tau$ (this is done by dropping singleton industries for both values of $\tau$). Figure \ref{fig4} A-F illustrates this idea. We compare the r-squared of CE$^\tau$ (for all $\tau$) to the r-squared of CE$^1$ (many small clusters), CE$^4$ and CE$^{10}$ (few large clusters). If, for example, CE$^{10}$ results in huge labour pool estimates due to overly large cluster sizes, then we would expect the r-squared corresponding to CE$^{10}$ to be smaller than more appropriate cluster definitions (values of $\tau$). In each case, we compute the difference in r-squared ($\Delta$R2) between the two versions of CE at each resolution $\tau$ in order to find values of $\tau$ which consistently yield larger r-squared scores (while holding the number of observations constant within each comparison).

We observe a peak in $\Delta$R2 for CE$^\tau$ relative to both CE$^1$ and CE$^{10}$ around $\tau=4$ (Figure \ref{fig4} B and F). We can interpret these results as suggesting that CE$^\tau$ is more strongly associated with employment growth than CE$^1$ or CE$^{10}$ for a range of $\tau$ around $2\leq\tau\leq5$. In contrast, when we compare CE$^\tau$ to CE$^4$ for all $\tau$ (Figure  \ref{fig4} D), at no point does CE$^\tau$ 'out-perform' CE$^4$. We infer from this that $\tau=4$ is the (statistically) optimal scale at which labour pooling occurs for services industries in the UK case.

In Figure \ref{fig4} G we repeat this exercise for all pairwise combinations of CE$^\tau$. Darker rows (and, symmetrically, columns) signify values of $\tau$ which tend to produce a larger values of r-squared relative to other resolutions. The results confirm what we saw above, with a peak around $\tau=4$.

\begin{figure*}[t!]
\centering
\includegraphics[width=\linewidth]{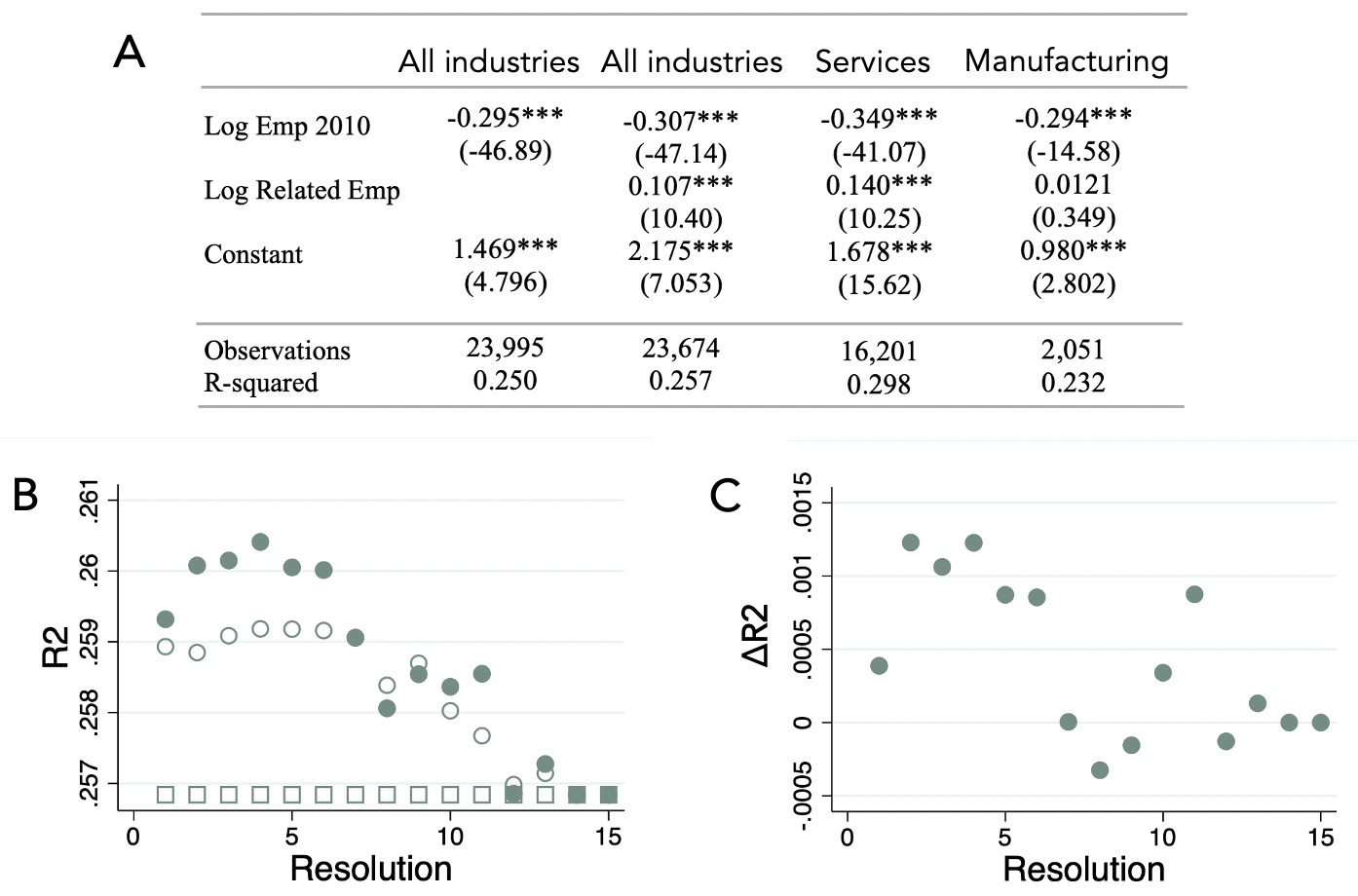}
\caption{A. OLS for industry employment growth as a function of log employment and related employment (RE) in the base year as defined by Eqn \ref{main} - for all industries, manufacturing only and services only. FE for cities and industries are included. B. We compare the r-squared of CE at each resolution with RE, including all observations (squares) and dropping observations such that CE and RE have the same observations (open circles). C. We show $\Delta$R2 as the difference in r-squared between CE and RE (same observations). \label{fig5}}
\end{figure*}

Finally, we compare cluster employment with its close cousin 'related employment' (RE). Related employment is essentially equivalent to the single community case of CE (i.e., resolution $\tau\geq 14$). Hence, related employment neglects the network structure, and does not capture the presence of industry clusters limiting skill-sharing. Consistent with the wider literature, Figure \ref{fig5} A shows that related employment is positively associated with employment growth for the set of all industries and services, but insignificant for manufacturing alone. 

Taking a similar approach as above, we compare the r-squared of CE$^\tau$ at each resolution with RE. In Figure \ref{fig5} B we show the r-squared for two models with RE, the first includes all observations (open squares) and the second drops observations such that CE and RE have the same observations (open circles). As above, we observe a clear difference in r-squared between the two models for a range of $\tau$ around $\tau=4$. In Figure \ref{fig5} C we compute $\Delta$R2 as the difference in r-squared between CE and RE with identical observations (closed and open circles respectively). These results lend further evidence to the idea that it is at these scales or resolutions at which clusters extracted from the network best explain industry employment growth, and do so better than a metric that implicitly ignores the network structure. 

In the Appendix, we provide a range of additional figures. These include robustness checks for variation in edge threshold parameter $\gamma$ and in the number of neighbours used in the construction of RE. Specifically: 
\begin{itemize}
    \item Figure \ref{app2} investigates robustness to changes in edge threshold $\gamma$. Here we repeat the analysis of Figure \ref{fig5} B for values of $\gamma=$0, 0.1, 0.3 and 0.5. We find that for a perturbation of the threshold up to 0.5, the analysis continues to exhibit similar behaviour.
    \item Figure \ref{app3} replicates Figure \ref{fig5} B but constructs RE using k nearest neighbours (nodes connected by k highest edges, similar to \citet{hausmann2021implied}). We find that the r-squared for CE peaks at a higher level than RE for all values of $k$.
\end{itemize}

\section{\bf{Discussion and Policy Implications}}
\label{sec:conclusionpolicy}

In this paper we extend the literature on defining industrial clusters ignited by \citet{porter1998clusters, porter2003economic} and later \citet{delgado2016defining}, and develop the conceptual and modelling toolbox of evolutionary economic geography by connecting commonly used metrics for 'related employment' to skill-based industrial clusters. In particular, we develop a methodology based on multi-scale community detection to uncover the statistically optimal scale at which inter-industry labour pooling operates in terms of supporting industry employment growth. Applying our model to data from the UK, we find that key sectors merge to form integrated labour pools. 

We contribute to a current push to exploit tools from network and data science to probe the complex dynamics underlying the labour market \citep{park2019global, moro2021universal}. Specifically, we link the topological structure of the labour mobility network to industry employment growth dynamics in a way that is both novel and founded in economic theory.   

\subsection{\bf{Policy Implications}}
\label{sec:policy}

Polices often suffer from a 'data deficit' concerning skills \citep{datadeficit}, particularly concerning the matching of skills training with employer needs. The application of tools from data science to labour market data, as we do here, has the potential to make a significant contribution to this issue. In this case, we illuminate in greater detail than was previously possible the complex inter-connection structure of industries based on skill overlap. Our methodology to extract a particular resolution at which labour pooling operates provides granular information on which industry combinations effectively exchange skills and know-how in the economy - and even more importantly - which do not. This is relevant for a wide range of policy domains, and particularly those aimed at promoting industrial growth and diversification, and developing targeted educational offerings that combine multi-sector skills that span traditional disciplinary boundaries. Specifically, education and training policies can be enhanced by understanding inter-industry linkages and mobility patterns, and particularly missing linkages which might be a target for new courses or apprenticeships.

Much policy discussion has been focused on interventions to increase labour market resilience in the presence of either short term shocks or longer term technological change. A key component of labour market resilience is the issue of the mobility of workers. More mobile workers, with more opportunities for transition between sectors, are better protected in the event of an economic shock. For example, \citet{diodato2014} find that intersectoral mobility is an important factor for regional recovery from an employment shock, and \citet{straulino2021res} find that UK cities with greater ability to reallocate skills between sectors were more resilient to the 2008 Financial Crisis. Our analysis of the modular structure of the labour network can provide a granular picture of both potential and realised worker movements in a shock, particular one affecting a limited number of sectors. It could be used to design policies that promote resilience, for example by creating new career pathways, or support workers with targeted schemes in the event of a crisis. 

Cities and regions benefit from a better understanding of their embedded skills and capabilities, particularly with respect to the development of evidence-based industrial policy. In recent years, insights from evolutionary economic geography have influenced the development of the EU's Smart Specialisation policy \citep{foray2014smart, boschma2014regional}. This policy advocates for the development of distinctive local clusters of businesses, building on a region's existing strengths. The methodology developed here is well-placed to further quantify the existence of particular skill-based clusters in a region, and uncover potential new industries which are closely related, requiring similar capabilities to those already present. 

\subsection{Limitations and Future Work}

On a methodological level, we implicitly assume that skill clusters are fixed over time. While this assumption is backed up by some analysis of similar German data \citep{neffke2017inter}, an interesting avenue of future research would be to investigate the patterns and drivers of change in the modular structure of the network over longer periods of time which is particularly pertinent for the design of longer term policies, such as higher level education. Similarly, short term periods of high mobility in response to a shock is of key interest, particularly in the case of Covid 19. 

Our study is limited by the nature of the data required to study worker inter-industry labour mobility. Longitudinal micro data on individual workers is only available in very limited circumstances, and is not even collected in all countries, even highly developed ones with advanced statistical capabilities. If collected, it is typically only available under limited access agreements. While efforts have been made to standardise and compare labour networks across countries \citep{straulino2021bi}, further work is required to build and release comparable administrative datasets for cross-country studies. 

Finally, while the idea that the presence of skills and capabilities in 'related' sectors benefits employment growth is well-established, the precise mechanisms by which firms access these skills and facilitate transitions remains less well understood. While a number of recent studies aim to more precisely measure the skill content of firms and industries using occupation and task based data (see e.g., \citet{neffke2019value, turco2020knowledge}), further investigation of worker traits and drivers of inter-industry mobility is warranted. 

\clearpage
\begin{footnotesize}
\bibliographystyle{elsarticle-harv}

\end{footnotesize}

\clearpage
\setcounter{section}{0}
\bf{\Large Appendix}

\section*{Data}

We use the Annual Survey of Hours and Earnings dataset (ASHE) in order to construct the inter-industry flow matrix. This is the most comprehensive source of earnings information in the United Kingdom to track workers job history, and contains anonymised demographic and employment information of 1\% of the total employee jobs in the HM Revenue \& Customs (HMRC) Pay As You Earn (PAYE) records, covering the years between 2009 to 2018 (before 2009 uses a different industry classification). For every worker, the dataset includes information on various variables, from individual characteristics, such as age and sex, to employment information, including pay, occupation and industry. The sample does not include the self-employed. 

We select workers between 16 and 65 years of age; and without missing information on industry, occupation, region of work, gender or age. For each individual, we identify the year of employment, the industry classification of each job and the post code of the work establishment associated to that job. Inter-industry flows correspond to changes in post code and industry code within a year.

For statistical disclosure reasons, only pairs of industries in which the total transitions summed across the entire period exceeds 10 observations are kept. In 2018, 3,142 flows were observed from 169,372 workers.

\section*{Community Detection}
\label{sec:community}
Random walks are a versatile dynamical system on graphs that can be used to study their structure. Intuitively, if we let a random walker wander on the network, and the walker remains in the same group of nodes over a long period of time, the group of nodes are tightly connected and represent a community. Here we review the Stability algorithm from \citet{delvenne2010stability}. 

Formally, we define the probability vector $P_{\tau} \in \mathbb{R}^n$ as the column vector which entry $i$ is the probability of finding a random walker at time $\tau$ in node $i$:
\begin{align*}
P_{\tau+1}=P_{\tau}D^{-1} A.
\end{align*} 
for adjacency matrix $A$ (and $D$ is a matrix of zeros with the node degrees on the diagonal). Observe that given an initial probability vector $P_0$, we have:
\begin{align*}
P_{\tau}=P_0 (D^{-1}A)^\tau. 
\end{align*}
We call the matrix $(D^{-1}A)^\tau$ the transition matrix at time $\tau$. In the case of non bipartite, non-directed and connected graph, for any starting point for the random walker, this process converges to a stationary probability distribution given by $\pi = d^T/(2 m)$. 

We will encode the partition in matrix $H \in \mathbb{R}^{n \times m}$, where $m$ is the number of communities in the partition, such that $H_{ij}=1$ if node $i$ is assigned for community $j$, and $0$ otherwise. We define the clustered auto covariance matrix of the diffusion process above as:
\begin{align}
\label{defR}
R(\tau,H)=H^T [\Pi P(\tau)- \pi^T \pi] H,
\end{align}
where $\Pi = \textrm{diag} (\pi)$. Observe that $(\Pi P(\tau))_{ij}$ represents the probability that the random walker who started from node $i$ ends up in node $j$ at time $\tau$. $(\pi^T \pi)_{ij}$ is the probability that the random walker, starting at node $i$, arrives at node $j$ at stationarity. Given our partition matrix, the diagonal entries of $R(\tau)$ therefore represent the probability for a random walker to remain in the community in which he started after $\tau$ has passed. We define the stability of a partition as:
\begin{align}
\label{defr}
r(\tau,H)=\textrm{Trace}(R(\tau)),
\end{align}
the sum of the diagonal elements of $R(\tau)$. As we want the walker to remain in the community in which he started, we seek a partition matrix $H$ that satisfies:
\begin{align}
\label{minrt}
H = \textrm{argmax}_{\hat{H}}r(\tau,\hat{H}),
\end{align}
on the set of all the possible partitions (for a given time $\tau$). 

This problem is NP-hard, and we therefore use heuristic methods, such as Louvain's algorithm \citet{blondel2008fast}, to solve it. This method first assigns each node to its own community. Then, for each node, it considers assigning it to a community with each of its neighbours one after another, and does so if it results in increased stability (given by the equation above). Once all nodes have been considered, nodes that were assigned to the same community are merged into one node, and the process is repeated until no increase in stability can be achieved. Observe that the result of algorithm depends on the order in which the nodes are considered.

Code to run the algorithm is available online at: \url{http://wwwf.imperial.ac.uk/~mpbara/Partition_Stability/}.

\section{Additional figures}

\renewcommand{\thefigure}{A\arabic{figure}}
\setcounter{figure}{0}

\label{sec:additional}
In this section a selection of additional figures to support the analysis of Section \ref{sec:results} are presented, including robustness checks for variation in edge threshold parameter $\gamma$ and in the number of neighbours used in the construction of RE. 

\begin{figure*}[t!]
\centering
\includegraphics[width=4in]{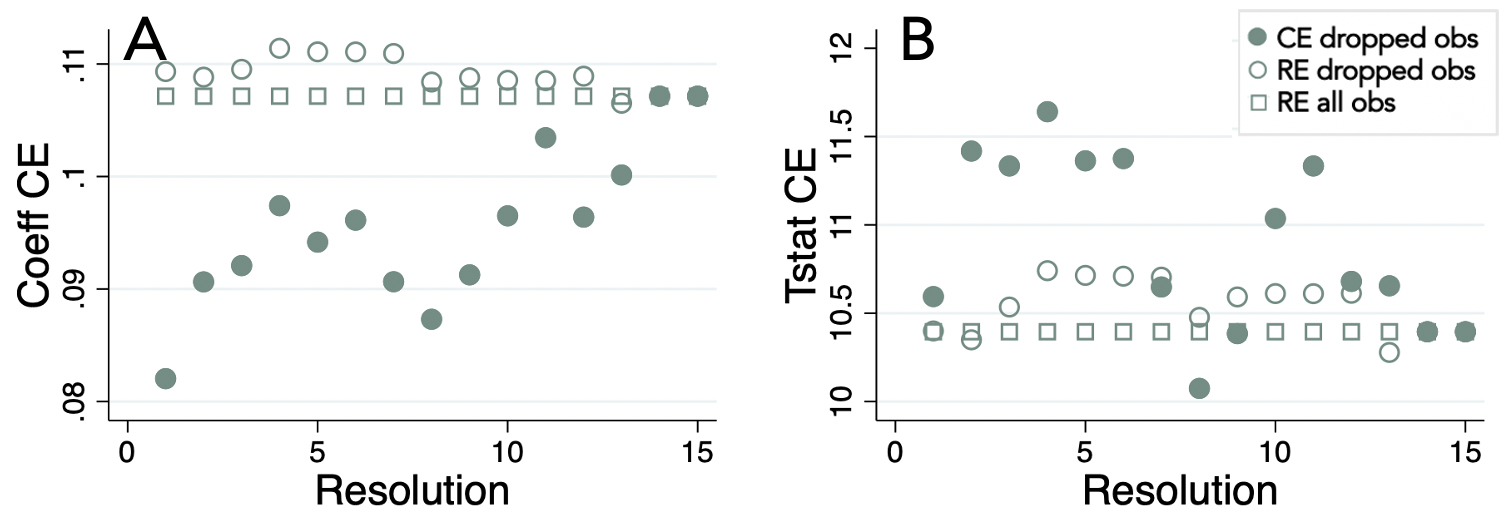}
\caption{The corresponding coefficient and t-statistic for Figure \ref{fig5} B. \label{app1}}
\end{figure*}

\begin{figure*}[t!]
\centering
\includegraphics[width=4in]{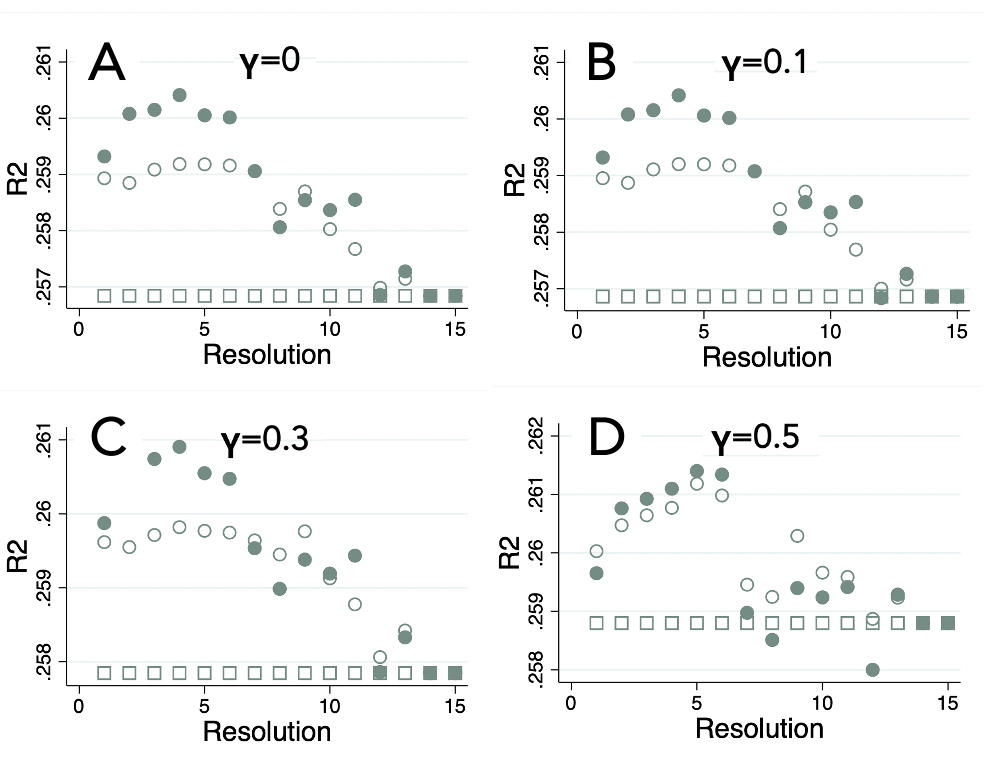}
\caption{Robustness to changes in edge threshold $\gamma$: here we repeat the analysis of Figure \ref{fig5} B for values of $\gamma=$0, 0.1, 0.3 and 0.5. For a  perturbation of the threshold ($\gamma=$0.1), we observe a consistent peak.\label{app2}}
\end{figure*}

\begin{figure*}[t!]
\centering
\includegraphics[width=6in]{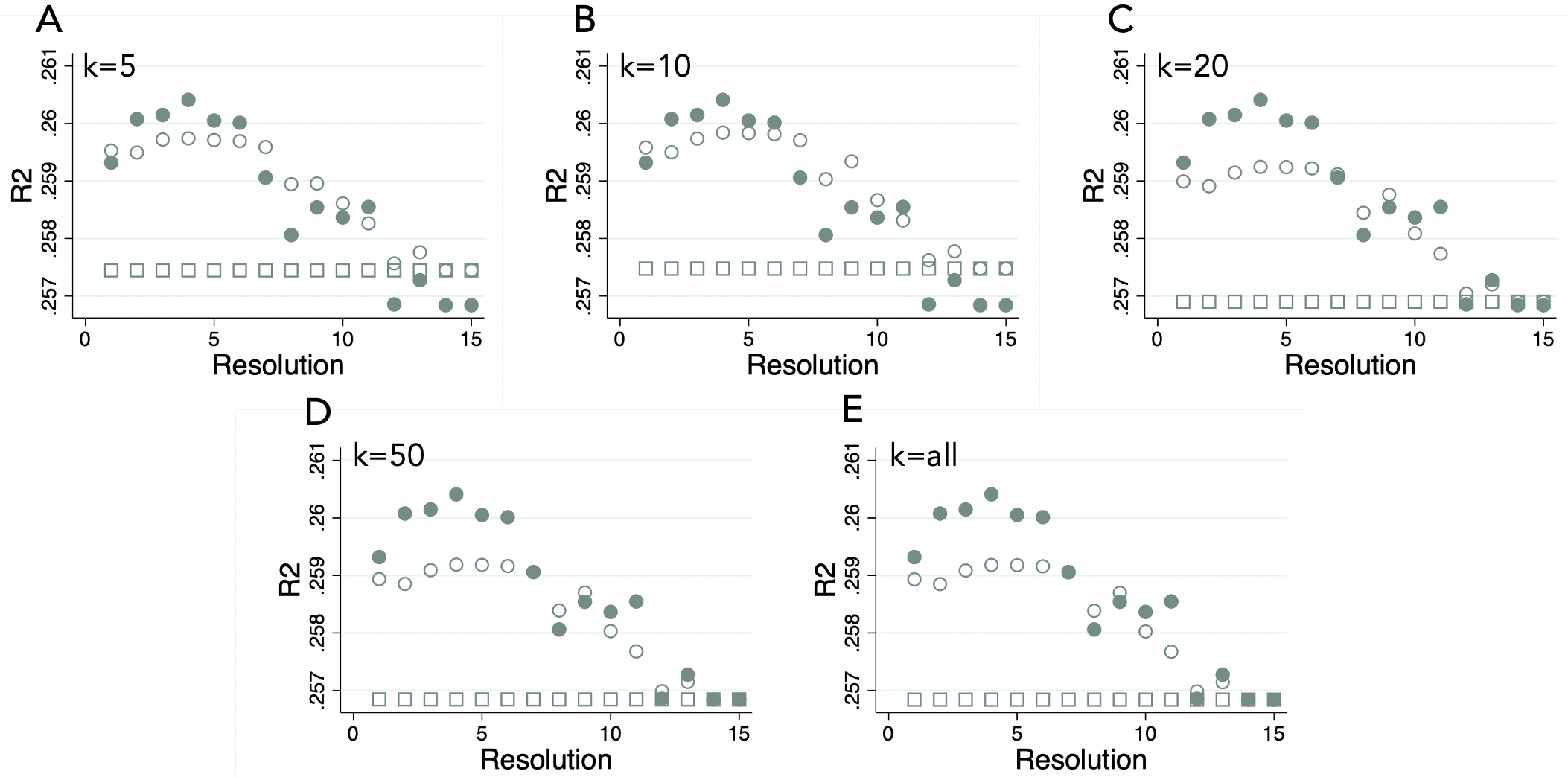}
\caption{Robustness to changes in number of neighbours in RE: this replicates Figure \ref{fig5} B but we construct RE using k nearest neighbours (we only include the k nodes connected by the k highest edge weights in the metric - k=all corresponds exactly to Figure \ref{fig5} B). We find that the r-squared for CE (filled circles) peaks at a higher level than RE (open circles) for all values of $k$.\label{app3}}
\end{figure*}

\end{document}